\documentclass[twocolumn,showpacs,preprintnumbers,amsmath,amssymb,footinbib,APS]{revtex4}

\usepackage{dcolumn}
\usepackage{bm}
\usepackage[spanish,english]{babel}
\usepackage{amsfonts}
\usepackage{amssymb}
\usepackage{graphicx}

\newcommand{\R}{\mathcal{R}}

\begin{document}

\title{ {\bf Violation of the Equivalence Principle in Modified Theories of Gravity}}
\author{Gonzalo J. Olmo\thanks{gonzalo.olmo@uwm.edu}}
\affiliation{ {\footnotesize Physics
Department, University of Wisconsin-Milwaukee, P.O.Box 413, Milwaukee, WI 53201,
USA }}


\date{December 31st, 2006}

\begin{abstract}
We study modified theories of gravity of the $f(R)$ type in
Palatini formalism. For a generic $f(R)$ lagrangian, we show that
the metric can be solved as the product of a scalar function times
a rank-two tensor (or auxiliary metric). The scalar function is
sensitive to the local energy-momentum density. The auxiliary
metric satisfies a set of equations very similar to Einstein's
equations and, for weak sources, it can be approximated by the
Minkowski metric. According to this, the metric coupled to the
matter strongly departs from the Minkowskian one in the
neighborhood of any microscopic physical system. As a consequence,
new gravitationally-induced interactions arise and lead to
observable effects at microscopic and macroscopic scales. In
particular, test body trajectories experience self-accelerations
which depend on the internal structure and composition of the
body. These facts make very unlikely the viability of Palatini
$f(R)$ models designed to change the late-time cosmic evolution.
\end{abstract}


\maketitle

Modified theories of gravity have been object of intense study in recent years due to their ability to predict cosmic acceleration \cite{Tonry-Knop} without the need for sources of dark energy. In this sense, special attention have received those theories of the $f(R)$ type, where $f$ is a non-linear function of the scalar curvature $R$, with nonlinear terms relevant at low (cosmic) curvatures. For a given lagrangian $f(R)$, such theories can be formulated in two inequivalent ways, namely, in metric and in Palatini formalism. In the former case, the connection is defined as the usual Levi-Civita connection. In the latter, the connection is regarded as independent of the metric and, therefore, must be determined by solving its corresponding field equations. According to this, the metric formalism leads to a system of fourth-order partial differential equations for the metric, whereas the Palatini formalism leads to second-order equations. Only for the linear lagrangian, $f(R)=R-2\Lambda$, the two formalisms lead to the same equations of motion, which are those of General Relativity (GR).\\
Though much has been written about the properties of $f(R)$ theories in cosmological applications (see \cite{CDTT04,metric} and \cite{Vol03,Palatini} for metric and Palatini formalism, respectively), we do not yet have a clear understanding of their properties in other regimes. Let us focus on Palatini theories. It is easy to verify (see below for details) that, for arbitrary $f(R)$, Palatini theories have the same type of vacuum equations as General Relativity with a cosmological constant. In particular, for a spherically symmetric star like the sun, the metric outside the star can be written as a Schwarzschild-de Sitter solution
\begin{equation}\label{eq:S-dS}
ds^2_{SdS}=g_{\mu\nu}dx^\mu dx^\nu=-A(r)dt^2+\frac{dr^2}{A(r)}+r^2d\Omega^2
\end{equation}
with $A(r)=1-2GM_\odot/r-\Lambda r^2/3$, where $\Lambda$, $G$ and $M_\odot$ represent a cosmological constant,   Newton's constant and the mass of the star respectively. The well-known model $f(R)=R-\mu^4/R$  of Carroll et al. \cite{CDTT04}, like any other $f(R)$ model, admits such solutions and leads to a tiny $\Lambda\sim\mu^2$. One is then tempted to conclude that this model, and any other model with a small $\Lambda$, is compatible with solar system observations \cite{Vol03,R-O06,A-R06} since, for sufficiently small $\Lambda$, (\ref{eq:S-dS}) is virtually undistinguishable from the Schwarzschild solution of GR, which passes all observational tests. However, though the vacuum solutions are very well known, the  equations in the presence of matter have not been studied in detail yet. The relation of the integration constant $M_\odot$ with the matter sources, for instance, is still unknown. And this is an important aspect, since Palatini theories only depart from GR in the presence of sources, as is well known in cosmology. Some results in this direction were found in the study of the Newtonian and/or post-Newtonian limits in \cite{Newtonian,Olmo05}. All those works agree, with small differences, in that in addition to the usual Newtonian potential $M/r$ the Newtonian limit has a density-dependent term. Such term was considered as irrelevant and negligible in some of those works. However, it was remarked in \cite{Olmo05} that it could be potentially dangerous. The reason being that it generates accelerations which depend on the gradient of the local matter density and which do not decay with distance. Furthermore, when an atomic/microscopic description of matter is considered, that density-dependent term seems to break the perturbative approach itself, which is a disturbing property. These facts suggest that further work is needed in order to fully understand the dynamics of Palatini theories.\\

In this work we study and solve the equations of motion of Palatini $f(R)$ theories in the presence of matter in some cases of interest, namely, spherically symmetric distributions and also for very weak sources without any particular symmetry. From our analysis it follows that strong gravitational effects arise at microscopic scales which are in conflict with the theoretical foundations of these theories and with the experimental evidence supporting the Einstein equivalence principle and our knowledge of elementary-particle physics. These facts strongly suggest that  $f(R)$ models designed to modify the gravitational interaction at low curvatures should be ruled out.\\

Let us begin by defining the action of Palatini theories
\begin{equation}\label{eq:Pal-Action}
S[{g},\Gamma ,\psi_m]=\frac{1}{2\kappa^2}\int d^4
x\sqrt{-{g}}f({R})+S_m[{g}_{\mu \nu},\psi_m]
\end{equation}
Here $f({R})$ is a function of ${R}\equiv{g}^{\mu \nu }R_{\mu \nu }(\Gamma )$, with $R_{\mu \nu }(\Gamma )$ given by
\begin{equation} \label{eq:Ricci}
R_{\mu\nu}(\Gamma )=-\partial_{\mu}
\Gamma^{\lambda}_{\lambda\nu}+\partial_{\lambda}
\Gamma^{\lambda}_{\mu\nu}+\Gamma^{\lambda}_{\mu\rho}\Gamma^{\rho}_{\nu\lambda}-\Gamma^{\lambda}_{\nu\rho}\Gamma^{\rho}_{\mu\lambda}
\end{equation}
where $\Gamma^\lambda _{\mu \nu }$ is the connection. The matter action $S_m$ depends on the matter fields $\psi_m$, the metric $g_{\mu\nu}$ and its derivatives, but not on $\Gamma^\lambda _{\mu \nu }$. Note that $\Gamma^\lambda _{\mu \nu }$ is regarded as a field of the gravity sector and, therefore, cannot appear in the matter sector, since only the metric is allowed to couple to matter according to the postulates of metric theories of gravity \cite{Will}. Varying (\ref{eq:Pal-Action}) with respect to the metric we obtain
\begin{equation}\label{eq:met-var-P}
f'(R)R_{\mu\nu}(\Gamma)-\frac{1}{2}f(R)g_{\mu\nu}=\kappa ^2T_{\mu
\nu }
\end{equation}
where $f'(R)\equiv df/dR$. Note that the trace of
(\ref{eq:met-var-P})
\begin{equation}\label{eq:trace-P}
f'(R)R-2f(R)=\kappa ^2T,
\end{equation}
implies an algebraic relation between $R$ and the
trace $T$. The solution to this algebraic equation will be denoted
by $R=\R(T)$. The variation of (\ref{eq:Pal-Action}) with respect to $\Gamma^\lambda _{\mu \nu }$ must vanish
independently of (\ref{eq:met-var-P}) and gives
\begin{equation}\label{eq:con-var}
\nabla_\rho  \left[\sqrt{-g}\left(\delta ^\rho _\lambda
f'g^{\mu \nu }-\frac{1}{2}\delta ^\mu _\lambda f'g^{\rho
\nu }-\frac{1}{2}\delta^\nu_\lambda f'g^{\mu
\rho}\right)\right]=0
\end{equation}
where $f'\equiv f'(\R(T))$ is also a function of the matter terms. Using
an auxiliary tensor $t_{\mu \nu }\equiv f'g_{\mu \nu }$,
(\ref{eq:con-var}) can be readily solved \cite{MTW73}. The solution
states the compatibility between $\Gamma^\lambda_{\mu \nu }$ and the {\it metric} $t_{\mu \nu}$.
In other words, $\Gamma^\lambda_{\mu \nu }$ can be written as
the Levi-Civita connection of $t_{\mu \nu}$
\begin{equation}\label{eq:Gamma-1}
\Gamma^\lambda_{\mu \nu }=\frac{t^{\lambda \rho
}}{2}\left(\partial_\mu t_{\rho \nu }+\partial_\nu
t_{\rho \mu }-\partial_\rho t_{\mu \nu }\right)
\end{equation}
Inserting this solution for $\Gamma^\lambda_{\mu \nu }$, written
in terms of $g_{\mu \nu }$ and $f'(\R(T))$, in
(\ref{eq:met-var-P}) we obtain
\begin{eqnarray}\label{eq:Gmn}
R_{\mu \nu }(g)-\frac{1}{2}g_{\mu \nu }R(g)&=&\frac{\kappa
^2}{f'}T_{\mu \nu }-\frac{\R f'-f}{2f'}g_{\mu \nu
}-\nonumber\\&-&\frac{3}{2(f')^2}\left[\partial_\mu f'\partial_\nu
f'-\frac{1}{2}g_{\mu \nu }(\partial f')^2\right]+\nonumber \\
&+& \frac{1}{f'}\left[\nabla_\mu \nabla_\nu f'-g_{\mu \nu }\Box
f'\right]
\end{eqnarray}
where $R_{\mu \nu }(g)$
and $R(g)$ are computed in terms of the
Levi-Civita connection of the metric $g_{\mu \nu }$, i.e., they
represent the usual Ricci tensor and scalar curvature. To make our
notation clearer, since $t_{\mu \nu }$ and $g_{\mu \nu }$ are
conformally related, it follows that $\R(T)=g^{\mu \nu }R_{\mu \nu
}(\Gamma)$ and $R(g)=g^{\mu \nu }R_{\mu \nu }(g)$ are related by
\begin{equation}
\R(T)=R(g)+\frac{3}{2f'(T)}\partial_\lambda f'(T)\partial^\lambda
f'(T)-\frac{3}{f'(T)}\Box f'(T)
\end{equation}
where, recall, $f'(T)\equiv f'(\R(T))$ is a function of $T$, which means that the $f'(T)$
terms in (\ref{eq:Gmn}) act as additional matter sources. The matter terms $\partial_\mu f'(T)$ and $\Box f'(T)$ make difficult the analisys of (\ref{eq:Gmn}) to obtain $g_{\mu\nu}$. In this sense,
we find very useful to shift the problem and find solutions for
the conformally related metric $\bar{g}_{\mu\nu}=\phi g_{\mu\nu}$, where $\phi\equiv f'(T)/f'(0)$ is a dimensionless factor which becomes unity in vacuum, $T=0$. The equations of motion for $\bar{g}_{\mu\nu}$ boil down to
\begin{equation}\label{eq:Gab-EF}
G_\mu^\nu(\bar{g})=\frac{\kappa^2}{f'_0\phi^2(T)}\left[T_\mu^\nu-\frac{V(T)}{2\kappa^2}\delta_\mu^\nu\right]
\end{equation}
where $f'_0\equiv f'(0)$ and we have used the shorthand notation $V(T)\equiv \R f'(T)-f(\R)$. The above equations of motion for $\bar{g}_{\mu\nu}$ are considerably simpler than those for $g_{\mu\nu}$, which
makes our task easier. In particular, for spherically symmetric systems we can use the ansatz $ds^2=g_{\mu \nu }dx^\mu dx^\nu=\phi^{-1}\bar{g}_{\mu \nu }dx^\mu dx^\nu$ with
\begin{equation}\label{eq:metric-int}
ds^2=\frac{1}{\phi}\left[-A(r)e^{2\Phi(r)}dt^2+\frac{1}{A(r)}dr^2+r^2d\Omega ^2\right]
\end{equation}
which leads to
\begin{eqnarray}\label{eq:Phi}
\frac{2}{r}\frac{d\Phi}{dr}&=&\frac{\kappa^2
}{f'_0\phi^2}\left(\frac{{T}_r^r-{T}_t^t}{A}\right) \\
-\frac{1}{r^2}\frac{d(r[1-A])}{dr}&=&
\frac{\kappa^2}{f'_0\phi^2}\left({T}_t^t-\frac{V(T)}{2\kappa^2}\right)
\label{eq:A}
\end{eqnarray}
Defining now $A(r)=1-2GM(r)/r$ in (\ref{eq:A}),
we can rewrite $M(r)$ and $\Phi(r)$ as
\begin{eqnarray}\label{eq:M}
M(r)&=&-\frac{\kappa^2}{2Gf'_0}\int_0^r dx \ x^2 \left[\frac{T_t^t-V(T)/(2\kappa^2)}{\phi^2}\right]\\
\Phi(r)&=&\frac{\kappa^2
}{2f'_0}\int^r_0dx \ x \left[\frac{{T}_r^r-{T}_t^t}{\phi^2A}\right]
\label{eq:Phi-1}
\end{eqnarray}
Outside of the sources, $(T_{\mu\nu}=0,\phi=1)$, the above equations can be readily integrated leading to
\begin{eqnarray}\label{eq:M-ext}
M(r)&=&M_\odot+\frac{V(0)r^3}{12Gf'_0}\\
\Phi(r)&=&\Phi_0 \label{eq:Phi-ext}
\end{eqnarray}
and we recover the Schwarzschild-de Sitter solution (\ref{eq:S-dS}), with $2\Lambda=V(0)/f'_0$ and $\Phi_0=$constant can be eliminated by a convenient redefinition of the time coordinate. Note that the constant $M_\odot$ must be fixed by matching the interior and exterior solutions. Now, since the interior solutions depend on the details of $T_\mu^\nu$ and the particular $f(R)$ lagrangian considered through $\phi(T)$ and $V(T)$, the value of $M_\odot$ (and also $\Phi_0$) will depend on the internal structure and composition of the object (for a recent example see \cite{KRS06}). In other words, a given amount of matter-energy, $\int dr r^2 T^t_t$, can lead to different external gravitational fields. This contrasts with General Relativity, where the external field generated by a spherically symmetric system only depends on the total matter-energy and not on the structure or composition (in GR $\phi=1,V(T)=0$). This behavior was already reported in \cite{Olmo05} within the post-Newtonian regime.\\

 We have just seen that outside of the sources (\ref{eq:metric-int}) turns into (\ref{eq:S-dS}). Let us consider what happens if we place a test body with $m\ll M_\odot$ within the external field of $M_\odot$. This test body could be anything from a laboratory-sized object to a single atom. The metric nearby this body is given by (\ref{eq:S-dS}). However, the metric inside of the test body, where $T\neq 0$, is given by (\ref{eq:metric-int}) with $\phi(T)\neq1$. In the $1/R$ model, for instance, we find $\phi=1-\frac{1}{2[1+\sqrt{1+12/\sigma^2}]}$, where $\sigma\equiv -T/T_c$ and $T_c=\mu^2/\kappa^2=\rho_\mu c$ with $\rho_\mu\sim 10^{-26} \ g/cm^3$. When $T$ drops below $T_c$ (outside the body) we have $\phi\to1$. For $T$ above $T_c$ (inside the body) we get $\phi\to 3/4$. We then see that the line element (\ref{eq:metric-int}) changes suddenly from $ds^2=ds^2_{SdS}$ to $ds^2=\frac{4}{3}ds^2_{SdS}$ when going from the outside to the inside of the test body. Actually, this happens in general when going from the outside to the inside of atoms, which leads to strong oscillations in the metric at microscopic scales both within the test body and within the central object characterized by $M_\odot$. This unusual and highly non-perturbative behavior, which has no precedent in the literature on alternative theories of gravity, is due to the dependence of the metric on the factor $\phi(T)$, which is a function of the local energy-momentum density $T$. Any theory in which the lagrangian $f(R)$ be sensitive to some low curvature scale will lead to a $\phi(T)$ characterized by two limiting values, $\phi\to 1$ for small $T$ and $\phi\to \phi_c$ for large $T$, as compared to the characteristic scale $T_c$. Therefore, the strong jumps in the metric discussed here for the $1/R$ model occur in all $f(R)$ models aimed at changing the late-time cosmic expansion.\\

To see in more detail the effects of the factor $\phi(T)$, we will
consider a portion of space-time containing the above test
body. Assume that we can take coordinates in which the line
element (\ref{eq:S-dS}) becomes Minkowskian away from the body,
i.e., $ds^2 \approx \eta_{\mu\nu}d\xi^\mu d\xi^\nu$. This choice
of coordinates eliminates the external field generated by
$M_\odot$. We could also have assumed no external field for
simplicity. The metric $g_{\mu\nu}$ in the region close to our
test body can then be computed using again the decomposition
$\bar{g}_{\mu\nu}=\phi g_{\mu\nu}$. For a sufficiently light test
body, such as an atom, the metric $\bar{g}_{\mu\nu}$ can be well
approximated by $\eta_{\mu\nu}$. This is so because
$\bar{g}_{\mu\nu}$ is governed by (\ref{eq:Gab-EF}), which is
dynamically very similar to General Relativity. In fact, the term
$\kappa^2/(f'_0\phi^2)$ can be seen as an effective (density
dependent) {Newton's constant}, and $V(T)$ as some (density
dependent) cosmological constant. For lagrangians of the form
$f(R)=R+\epsilon g(R)$, with $\epsilon$ some small parameter, we
find that $ V(T)\sim \epsilon$. Thus $V(T)$ is very small (of
cosmological order) and can be neglected in the region of
interest. Like in General Relativity, the contribution of
$T_\mu^\nu$ is also negligible for weak sources. Therefore,
(\ref{eq:Gab-EF}) becomes $G_\mu^\nu (\bar{g})\approx 0$ and leads
to  $\bar{g}_{\mu\nu}\approx\eta_{\mu\nu}$. The metric generated
by our test body (atom) is then
\begin{equation}\label{eq:phi-eta}
g_{\mu\nu}(x)=\phi^{-1}[T(x)]\eta_{\mu\nu}
\end{equation}
Note that the factor $\phi^{-1}[T(x)]$ is due to the derivatives of $f'(T)$ on the right hand side of (\ref{eq:Gmn}) which, unlike $T_{\mu\nu}$ and $V(T)$, cannot be neglected. We thus see that the metric in the neighborhood of any small physical system, described by a certain probability density distribution, is not in general the Minkowski metric. This result is very important and leads to a number of inconsistencies and pathological effects, which must be added to those discussed above.\\
Let us first focus on the theoretical inconsistencies. The action (\ref{eq:Pal-Action}) was constructed according to the postulates of metric theories of gravity, namely, i) spacetime is endowed with a symmetric metric $g_{\mu\nu}$, ii) the world lines of test bodies are geodesics of that metric, and iii) in local freely falling frames the nongravitational laws of physics are those of special relativity. These postulates tell us how to introduce matter in a theory of gravity based on geometry: we must take the Minkowskian theory and change $\eta_{\mu\nu}$ by $g_{\mu\nu}$ and $\partial_\mu$ by $\nabla_\mu$, with $\nabla_\alpha g_{\mu\nu}=0$. This prescription should guarantee that the non-gravitational laws of physics of Minkowski space were recovered in local freely falling frames. In our case, however, even in local frames in which external gravitational fields have been screened, we do not recover the Minkowskian metric, which violates postulate iii).\\
On the other hand, since these postulates assume that the Einstein Equivalence Principle (EEP) is valid, it follows that Palatini $f(R)$ theories must be in conflict with the experimental evidence supporting the EEP (see \cite{Will} for details). Let us thus focus on the weak equivalence principle (also known as universality of free fall, and a key piece of the EEP), which states that the trajectory of a freely falling test body in a local inertial frame is unaccelerated and independent of the internal structure and composition of the body. From (\ref{eq:phi-eta}) we see that the geodesic equation
\begin{equation}\label{eq:geodesic}
\frac{d^2\xi^\mu}{d\tau^2}+C^\mu_{\alpha\beta}(g) \frac{d\xi^\alpha}{d\tau}\frac{d\xi^\beta}{d\tau}=0,
\end{equation}
where $C^\mu_{\alpha\beta}(g)=\frac{g^{\mu \rho}}{2}\left(\partial_\alpha g_{\rho \beta }+\partial_\beta
g_{\rho \alpha }-\partial_\rho g_{\alpha\beta}\right)$, leads to
\begin{equation}\label{eq:geodesic-t}
\frac{dv^i}{dt}=\frac{1}{2}(1-|\vec{v}|^2)\left[v^i\partial_t \ln \phi +\partial^i \ln \phi\right]
\end{equation}
where $v^i=d\xi^i/dt$. This equation shows that if $\partial_\mu \phi(T)\neq 0$, the test body will feel self-accelerations which, in addition, will depend on its own internal structure and composition.  Now, the deep dependence of $\phi(T)$ on the microscopic structure does not allow us to use an averaged description of matter (perfect fluid) to study the behavior of (\ref{eq:geodesic-t}). One should use a microscopic description for the matter sources, as pointed out in \cite{Fla04a}. Let us then consider the action of a free Dirac field in Minkowski space and use the EEP to construct the corresponding curved space theory to be put in (\ref{eq:Pal-Action}). The result is
\begin{equation}\label{eq:Dirac-cs}
S_m=\int d^4x \sqrt{-g}\left[\bar{\psi}\gamma^\mu\nabla_\mu \psi+im\bar{\psi}\psi\right],
\end{equation}
where $\nabla_\mu\psi=(\partial_\mu-\Gamma_\mu)\psi$, $\Gamma_\mu$ is the spin connection, and $\{\gamma^\mu,\gamma^\nu\}=2g^{\mu\nu}$ are the curved-space counterparts of Dirac's constant matrices $\{\gamma^a,\gamma^b\}=2\eta^{ab}$. Though the Minkowskian theory should be recovered in a freely falling frame, where external gravitational fields have been screened, from  (\ref{eq:phi-eta}) and (\ref{eq:Dirac-cs}) we find
\begin{eqnarray}\label{eq:Dirac-fff}
S_m&=&\int d^4\xi\phi^{-2}(T)\times\\
&\times&\left[\phi^{1/2}(T)\bar{\psi}\gamma^a\partial_a \psi-\frac{3}{2}\partial_a \phi^{1/2}(T)\bar{\psi}\gamma^a\psi+im\bar{\psi}\psi\right]\nonumber
\end{eqnarray}
where $T$ depends on all the sources present, including $\psi$ itself, and $\partial_a \phi^{1/2}$ is due to $\Gamma_\mu$ \footnote{Note that $\phi(T)$ can be obtained from (\ref{eq:trace-P}) without solving the equations of motion for the matter fields.}. It is quite evident that (\ref{eq:Dirac-fff}) is not the Minkowski space theory we started with and that new gravitationally-induced (first-order) interactions arise \footnote{This result is independent of the choice of conformal frame for the gravitational action} via the various $\phi(T)$ terms in (\ref{eq:Dirac-fff}). This modification of the microscopic interactions is what eventually leads to the macroscopic violations of the universality of free fall observed in (\ref{eq:geodesic-t}). Note that this effect is different in nature from the {\it Nordtvedt effect} \cite{Nordtvedt}, which is due to the coupling of the gravitational self-energy of a massive body to external gravitational fields (second order gravity-gravity interaction). On the other hand, the fact that $\phi(T)$, in models relevant for the late-time cosmology, is sensitive to low energy-density scales, makes it difficult (or perhaps impossible) an expansion of (\ref{eq:Dirac-fff}) in powers of $T$ to quantitatively study the  predictions of particular models. This means that the non-gravitational laws of physics of Minkowski space are strongly modified and might even lose their predictive power. This problem also affects the gravitational laws, since the metric is sensitive to the microscopic structure of $T_{\mu\nu}$, and we must solve the microscopic matter field equations to get $T_{\mu\nu}$. This fact, by the way, invalidates all cosmological applications of Palatini theories considered so far \cite{Vol03,Palatini}, in which an averaged description of matter (perfect fluid) has been assumed.\\
In summary, we have shown that the gravitational sector in Palatini $f(R)$ theories has dramatic effects on the matter sector. When the gravitational interaction is turned on, the Minkowskian laws of physics are completely lost due to the sensitivity of the spacetime metric to the local energy-momentum densities. This fact, besides being a theoretical inconsistency, implies observable effects such as violations of the EEP and new interactions among elementary particles. Furthermore, if the theory is sensitive to very low density scales, it might be impossible to use the theory to extract quantitative predictions (see \cite{Fla04b} for an example). All this suggests that Palatini $f(R)$ theories, at least in their current form, should be ruled out. Only those models in which $\phi(T)$ is  only sensitive to very high energy densities, above the currently accessible experimental limits, might have a chance to survive.\\

This work has been supported by NSF grants PHY-0071044 and PHY-0503366.

\end{document}